# The seismogenic area in the lithosphere considered as an "Open Physical System". Its implications on some seismological aspects. Part - III. Seismic Potential.

Thanassoulas[1], C., Klentos[2], V.

1. Retired from the Institute for Geology and Mineral Exploration (IGME), Geophysical Department, Athens, Greece.
   e-mail: thandin@otenet.gr - website: www.earthquakeprediction.gr

2. Athens Water Supply & Sewerage Company (EYDAP),
   e-mail: klenvas@mycosmos.gr - website: www.earthquakeprediction.gr

**Abstract.**

The seismic potential of any regional seismogenic area is analyzed in terms of the "open physical system" inflow – outflow energy balance model (Thanassoulas, 2008, Part – I). Following the magnitude determination method presented by Thanassoulas, (2008, Part – II) any region of any arbitrary area extent is assumed as being a potential seismogenic region. Consequently, the capability for the generation of a maximum magnitude future EQ at each virtual seismogenic region is investigated all over Greece at certain times. The later results are used to compile maps of the seismic potential / maximum expected EQ magnitude for Greece at 5 year's intervals ranging from 1970 to 2000. The comparison of these seismic potential maps / maximum expected EQ magnitude to the corresponding seismicity (M>6R) for each corresponding 5 years period reveals their tight interrelation. Therefore, the calculated seismic potential / maximum expected EQ magnitude, due to its drastic change in time in any seismogenic region, is a dynamic in time parameter which indicates the seismic energy charge status of each seismogenic area.

## 1. Introduction.

The terms "seismic hazard" and "seismic risk" are, very often, referred to in the seismological and engineering geology studies.

The term "seismic hazard", at any place, refers to a quantity **(H)**, its magnitude being the expected intensity of the ground motion at this place. The later, can be expressed as (Papazachos et al 1985, 1989, Tselentis 1997) the expected ground acceleration, ground velocity, ground dislocation and the expected, macroseismic intensity **(I)**.

The term "seismic risk" **(R)** refers to the expected results (damages in buildings, deaths etc) from the occurrence of an earthquake and depends strongly on the seismic hazard of the same place. The term **(R)** of the seismic risk can be expressed as the convolution of the seismic hazard **(H)** to the vulnerability **(V)** of a technical construction. Therefore, the following equation holds:

$$R = H * V \tag{1}$$

Tselentis (1997) presents the following, holding equation for the seismic risk:

$$R = H(e, \mu, s) * T \tag{2}$$

Where, **(R)** is the seismic risk, **(H)** is a non-linear parametric **(e, μ, s)** equation with **(e)** being the earthquake source parameters, **(μ)** is the propagating elastic waves media, **(s)** is the local conditions and **(T)** is the vulnerability of the technical constructions.

A seismic risk study, at a certain place, has a strong probabilistic – stochastic character and therefore all parameters that can contribute to an excess ground motion at a probabilistic level, are taken into account.

The results of a seismic risk study are presented in various forms. Probabilistic graphs vs. Mercalli scale, excess of spectral velocity and maps of spatial distribution of expected ground velocity are some of them.

A typical, seismic risk study of a place includes (Tselentis, 1997) the four following basic steps:

- Identification of the near-by earthquake sources.

- Determination of the statistical model that prescribes the earthquake sources and the expected, maximum magnitude due to each one of them.

- Determination of the best amplitude decay of the seismic waves of each seismic source.

- Determination of the probability for non-exceeding any ground motion parameter level.

Maps of the spatial distribution of seismic hazard of Greece were presented in the past (Makropoulos et al. 1985, Papazachos et al. 1985, 1989). Furthermore, the Greek territory was divided in four **(I, II, III, IV)** zones of different expected ground acceleration, as a function of the recurrence mean time value, and the intensity **(I)** of a future, seismic event. This particular former seismic hazard map of Greece is presented in the following figure **(1)**.

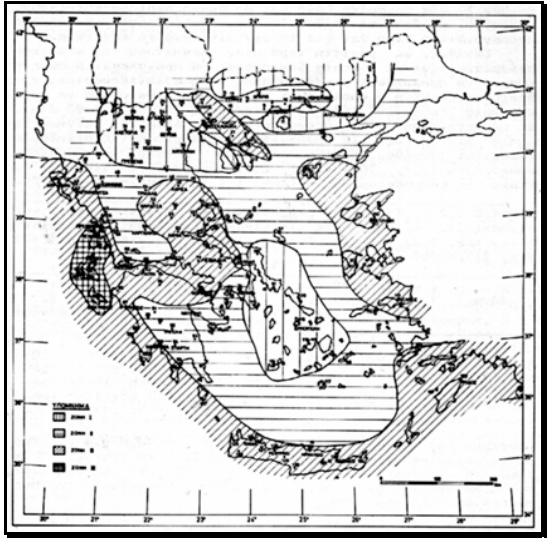

Fig. 1. Former, seismic hazard zoning is presented for the Greek territory (Papazachos et al. 1989, OASP).

This map has already being revised, due to the large seismic events that took place in Greece, during the last decade (1990 – 2000). The new map is divided into three zones and presents a close resemblance to the one of figure (**1**).

Following the mathematical analysis which is presented by Papazachos et al. (1989), it is made clear that the seismic hazard map of Greece is based mainly on probabilistic seismic data, as far as it concerns the parameters of the earthquake sources.

The later was verified, when Kozani earthquake (6.6R, 13/05/1995) and Athens earthquake (5.9R, 7/9/1989) took place at areas, which were considered before as, more or less, aseismic.

Therefore, that map should be modified, as soon as possible, whenever new, seismic data are available from earthquakes that have already occurred, or even better, it should be modified according to already known data, from areas that are being highly charged with strain energy and consequently strong earthquakes are expected to occur, within some period of time (a few years).

In the recent decade, a seismological research trend was developed towards earthquake prediction (medium term prediction) by the use of accelerating seismic energy release or accelerated deformation, as it is very often referred to. Main (1995), studied the earthquakes from the point of view of a critical phenomenon. Varnes (1987a, b, 1989), related the released, seismic energy to earthquake foreshock sequences in an attempt to predict earthquakes by analyzing accelerating, precursory seismic activity. A simple, in time **(t)**, empirical power law failure function was postulated by Bufe and Varnes, (1993) that relates the parameters of the remaining time **($t_c$-t)** to the occurrence of the imminent strong EQ and its corresponding magnitude **(M)** to the seismic moment release. In the same area of Statistical Physics, Main (1996) showed that the cumulative, seismic strain release increases as a power law time to failure before the final event. Bowman et al. (1998) used the concept of cumulative, seismic strain release that increases as a power law time to failure, before the final event. Moreover it was found that the critical region of radius **(R)** and the magnitude of the final event (M) are correlated as: **Log(R) = 0.5M,** suggesting that the strongest probable event, in a given region, scales with the size of the regional fault network.

Papazachos et al. (2000, 2001, 2001a, 2002), applied the same methodology to the Aegean area, Greece. In this case, ellipses were taken into account for the critical regions, which correspond to circles of radius **(R)** with equal area. Di Giovambatista et al. (2001) studied the accumulated Benioff strain before strong earthquakes, by using the time-to-failure model, and presented examples from strong earthquakes that occurred at the Kamchatka and in Italy. Tzanis et al. (2003) related the crustal deformation in SW Hellenic ARC to the distributed power-law seismicity changes.

Following this methodology, it is obvious, theoretically, that areas of increased probability, for the occurrence of a strong EQ, can be identified in advance and therefore, the expected, maximum magnitude of an imminent EQ, at any place, will affect, accordingly, the seismic hazard, calculated, for it.

Such maps, prepared for the entire Greek territory, will modify, accordingly, the seismic hazard map, proposed by the seismologists, which are, already, in use by the state authorities.

The above methodology has a main drawback. The acceleration of seismic energy release is not a universal process that is observed before all strong earthquakes. The application of the time-to-failure model, for the estimation of the magnitude and moment of a future, strong earthquake, is possible only in case when a tendency to acceleration is observed in the release of seismic energy in the vicinity of its epicenter. Moreover, it does not distinguish between the earthquake swarm and the foreshock activation.

This work presents a different approach in preparing such maps. These maps can be compiled, in particular, by the application, all over the Greek territory, of the energy flow model of the lithosphere (Thanassoulas et al. 2001, Thanassoulas 2007, 2008a, b) on the past seismic data.

Particularly, any area of any extent can be considered as a virtual seismogenic area. Therefore, its past seismic history can be studied. Consequently, the stored seismic energy in this region can be calculated following the method presented by Thanassoulas (2008, Part – II). This operation can be repeated on a regular grid over a seismogenic country (i.e. Greece) and the obtained results can be used for the compilation of maps for the spatial distribution of the seismic charge / seismic potential (in terms of expected maximum magnitude of a future EQ) of it. The operation can be repeated in different times so that the change in time of the seismic potential / expected maximum magnitude of a future EQ can be evaluated.

## 2. The data.

The used seismic data were downloaded from the National Observatory of Athens, Institute of Geodynamics, (NOA), Greece.

Greece is, initially, entirely considered as a unit critical seismogenic region and the lithospheric energy flow model was applied over it.

The cumulative energy **(Ec)** vs. time graph **(fig. 2)**, for the period 1950 – 2002, reveals the presence of several **(A, B, C)** linear parts of it, with different energy flow rate values **(EFLs)**. These linear parts are intermitted by periods **(1, 2, 3,** and **4)** of intense, seismic activity.

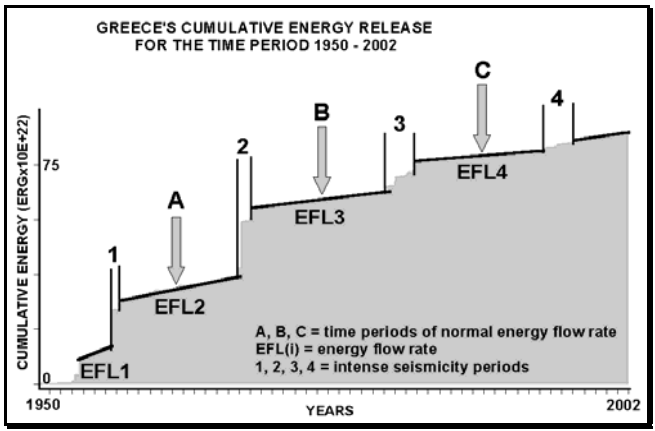

Fig. 2. Greece's cumulative energy release is presented for the time period 1950 – 2002 (Thanassoulas et al., 2003).

The continuous decrease of the **EFL** values (energy flow rate), which is observed from 1950 to 2002, indicates that for the last 52 years, Greece, was charged continuously with seismic strain energy that will be released some time in the future, at some seismically prone areas.
Consequently, two questions must be answered:

- Which areas have been charged, in excess, with seismic strain?

- What is the expected maximum value of a future earthquake, for the next i.e. five year's period, for each area?

The following procedure was followed, in order to answer these two questions.

### 3. Application of the method – examples.

The lithospheric energy flow model was applied over the Greek territory, by using a grid shell of 100 x 100Km. That shell slides all over Greece, in steps of 50Km from West to East and from North to South. As a result of this operation, an overlapping of 50% was applied on the obtained results and a final grid of 50 x 50Km shell was resulted.
This procedure is presented in the following figure **(3)**.

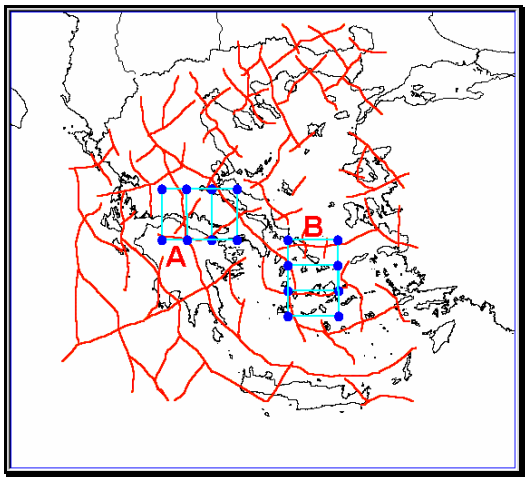

Fig. 3. Size of grid shell (blue frame) is shown and its movement over the Greek territory. **A =** shell movement from West to East. **B =** shell movement from North to South. Both movements are performed at steps of 50Km.

For each shell position and for each defined area, the corresponding, cumulative energy vs. time function, for the time period 1950 - 2000 was calculated. Determinations were made for the seismic energy strain charge potential, in terms of expected maximum magnitude, for a future earthquake, by using this graph and following the lithospheric energy flow model procedure. As a time basis for these calculations the years 1970, 1975, 1980, 1985, 1990, 1995, and 2000 were adopted, while the expected magnitude was calculated for each next five years period. For example, the calculation for the year 1980 indicates the maximum, expected magnitude for a strong earthquake in the period of 1980 – 1985.
The used shells total to a number of **(323)** and the obtained maximum expected magnitudes for the different time periods were used to compile the corresponding maps. These maps present the strain charge status spatial distribution all over Greece, in terms of maximum magnitude of a potential, future (within the next five years from the date of the map) earthquake.
Following, are the compiled maps for the corresponding periods of time. These maps are presented in two forms:

- The first one presents the spatial distribution of the entire range of the expected maximum magnitude of a possible, future earthquake over the Greek territory.
- The second presents the same results, as above, but with a low threshold level of magnitude set at 6 R. This facilitates to identify more easily the areas which are highly strain-charged and therefore, prone to intense, seismic activity.

In each map, a chromatic bar indicates the corresponding expected earthquake magnitudes in Richter scale. The compiled maps are presented in the next pages as follows:

**Year: 1970**

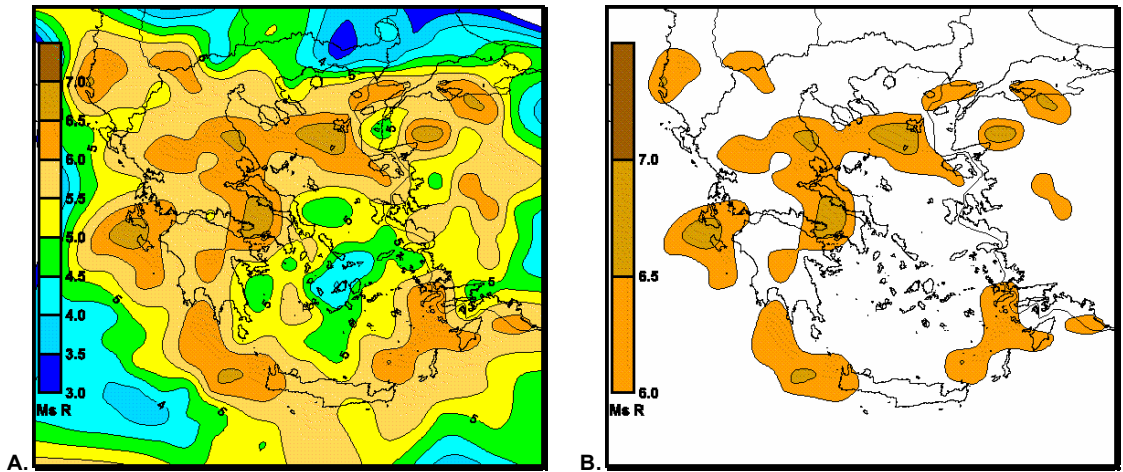

Fig. 4-5. **A**: compiled map based on data from 1950 to 1970 and expected potential earthquake magnitudes up to year 1975. **B**: the same as **A,** but a threshold of 6R was used.

**Year: 1975**

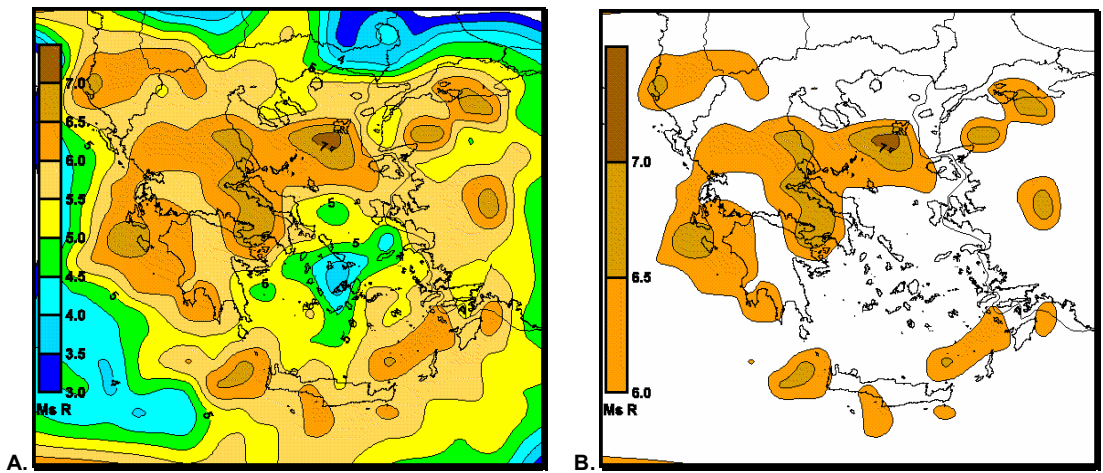

Fig. 6-7. **A**: compiled map based on data from 1950 to 1975 and expected potential earthquake magnitudes up to year 1980. **B**: the same as **A,** but a threshold of 6R was used.

**Year: 1980**

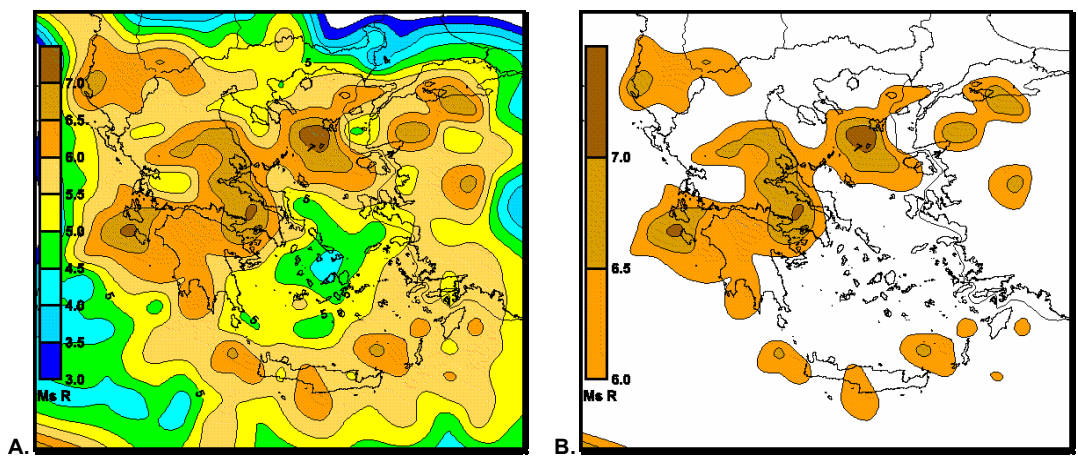

Fig. 8-9. **A**: compiled map based on data from 1950 to 1980 and expected potential earthquake magnitudes up to year 1985. **B**: the same as **A,** but a threshold of 6R was used.

**Year: 1985**

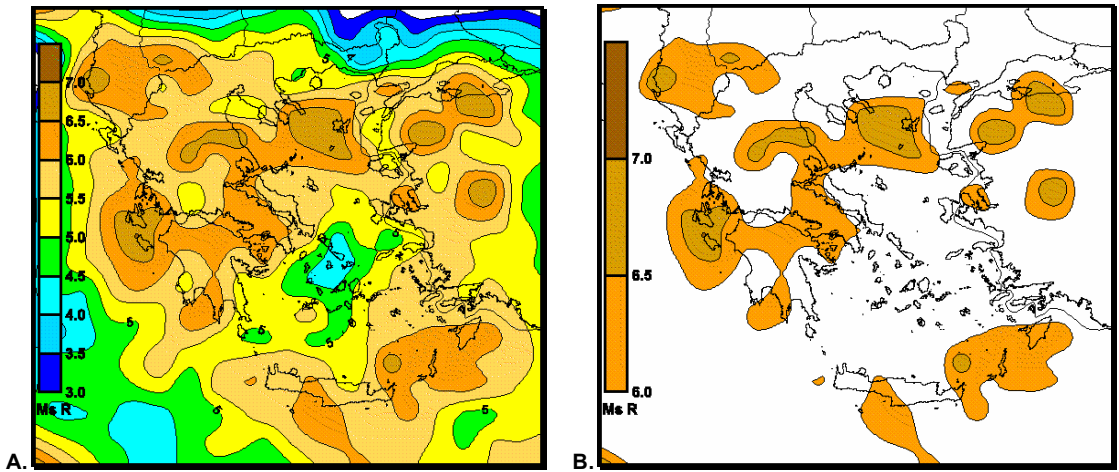

Fig. 10-11. **A**: compiled map based on data from 1950 to 1985 and expected potential earthquake magnitudes up to year 1990.
**B**: the same as **A,** but a threshold of 6R was used.

**Year: 1990**

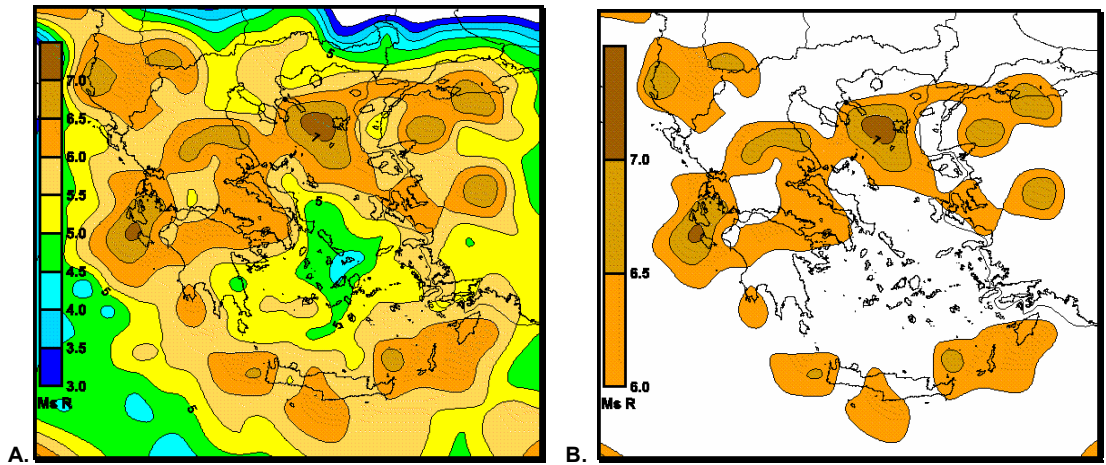

Fig. 12-13. **A**: compiled map based on data from 1950 to 1990 and expected potential earthquake magnitudes up to year 1995.
**B**: the same as **A**, but a threshold of 6R was used.

**Year: 1995**

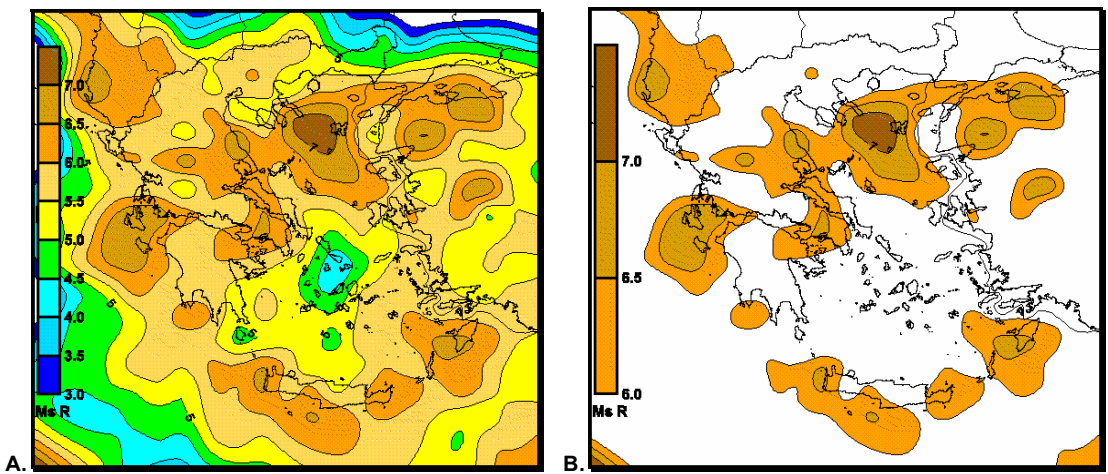

Fig. 14-15. **A**: compiled map based on data from 1950 to 1995 and expected potential earthquake magnitudes up to year 2000.
**B**: the same as **A**, but a threshold of 6R was used.

**Year: 2000**

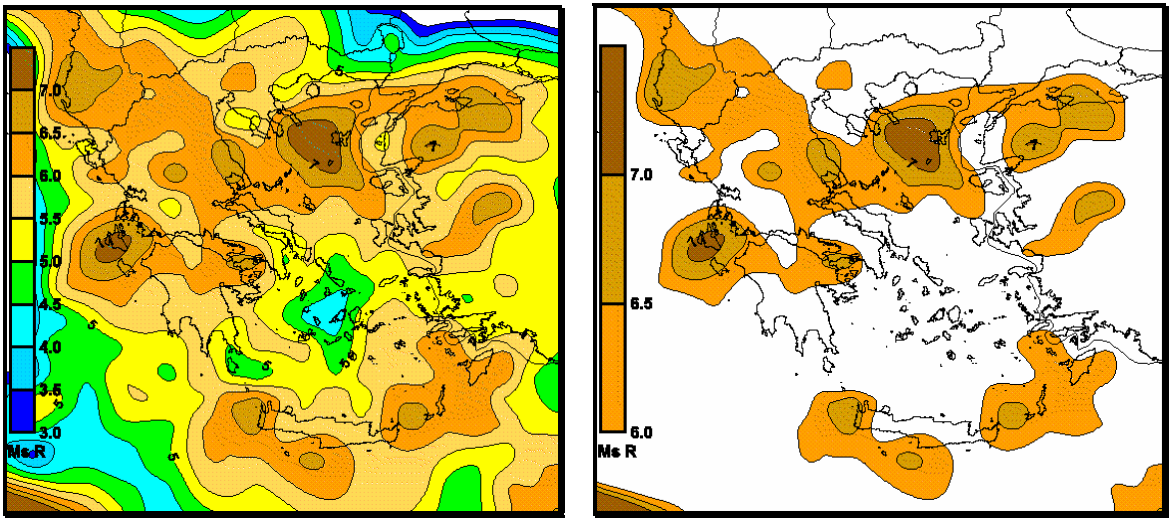

Fig. 16-17. **A**: compiled map based on data from 1950 to 2000 and expected potential earthquake magnitudes up to year 2005.
**B**: the same as **A,** but a threshold of 6R was used.

The subsequent maps are presented in order to validate the used method and the results which are obtained through it. On each map, which represents a specific period of time, the corresponding strong earthquakes (Ms >= 6R) which occurred during the next five years were superimposed. The percentage of the total earthquakes that occurred, in the predefined area (Ms>=6R), is calculated as a measure of the success of the method. That is the ratio of:

**P = EQin / EQtot**  (3)

Where **P** is the success percentage, **EQin** is the number of earthquakes that occurred in the predefined area and **EQtot** is the total number of the strong EQs that occurred in the specific period of time in the Greek territory.

**Period: 1970 – 1975**

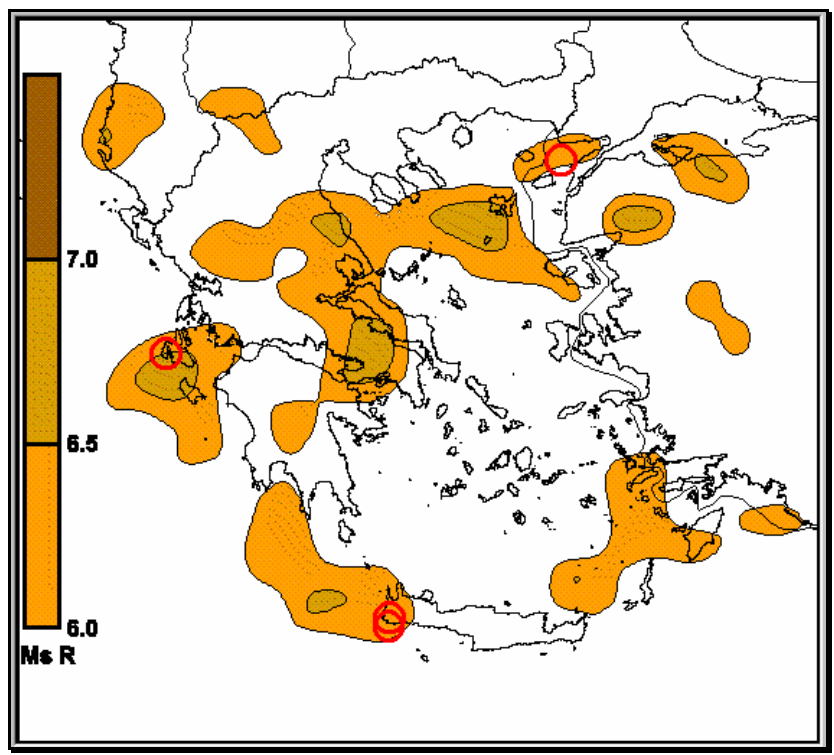

Fig. 18. Strong seismic events (red circles, Ms> = 6R) during the period 1970 – 1975.
**EQtot = 4, EQin = 4,  P = EQin / EQtot =  100%**

**Period: 1975 – 1980**

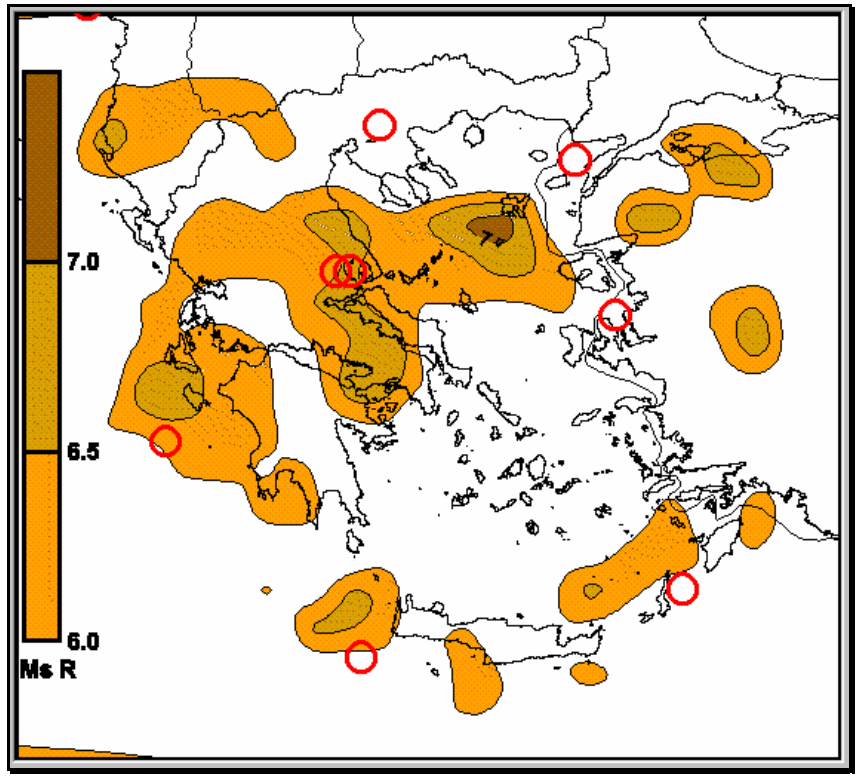

Fig. 19. Strong seismic events (red circles, Ms>=6R) during the period 1975 –1980.
**EQtot = 8, EQin = 5, P = EQin / EQtot = 62.5%**

**Period: 1980 – 1985**

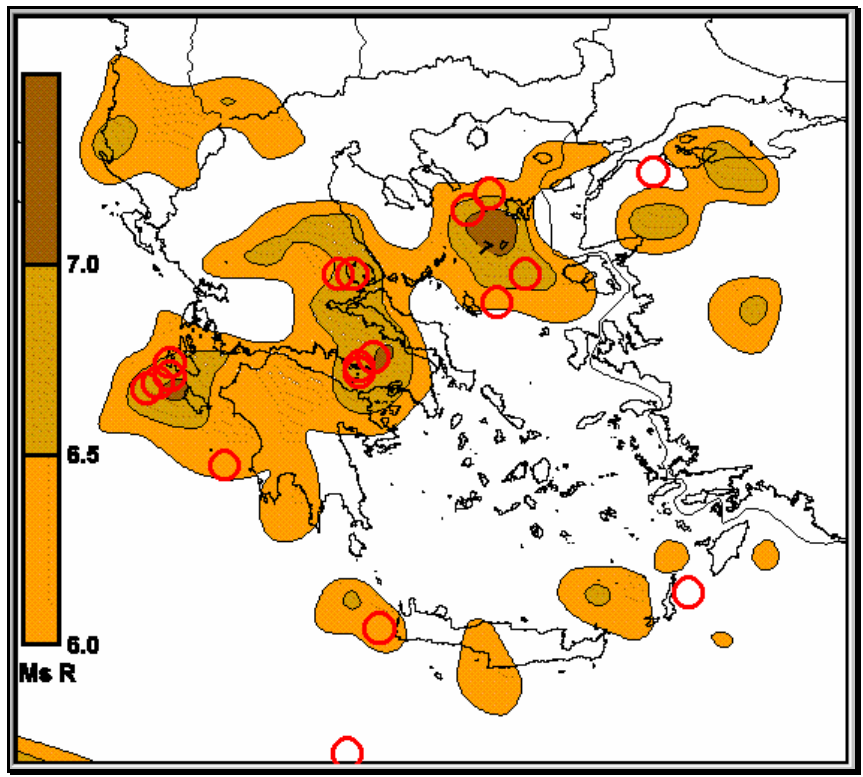

Fig. 20. Strong seismic events (red circles, Ms>=6R) during the period 1980 – 1985.
**EQtot = 17, EQin = 15, P = EQin / EQtot = 88.2%**

**Period: 1985 – 1990**

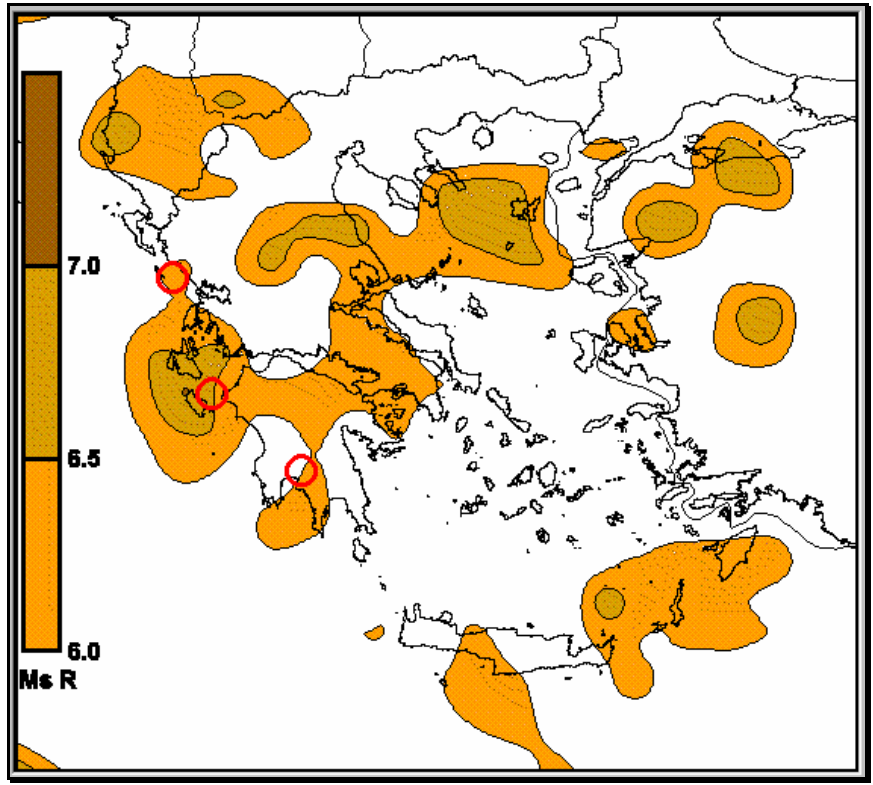

Fig. 21. Strong seismic events (red circles, Ms>=6R) during the period 1985 – 1990.
**EQtot = 3, EQin = 3, P = EQin / EQtot = 100%**

**Period: 1990 - 1995**

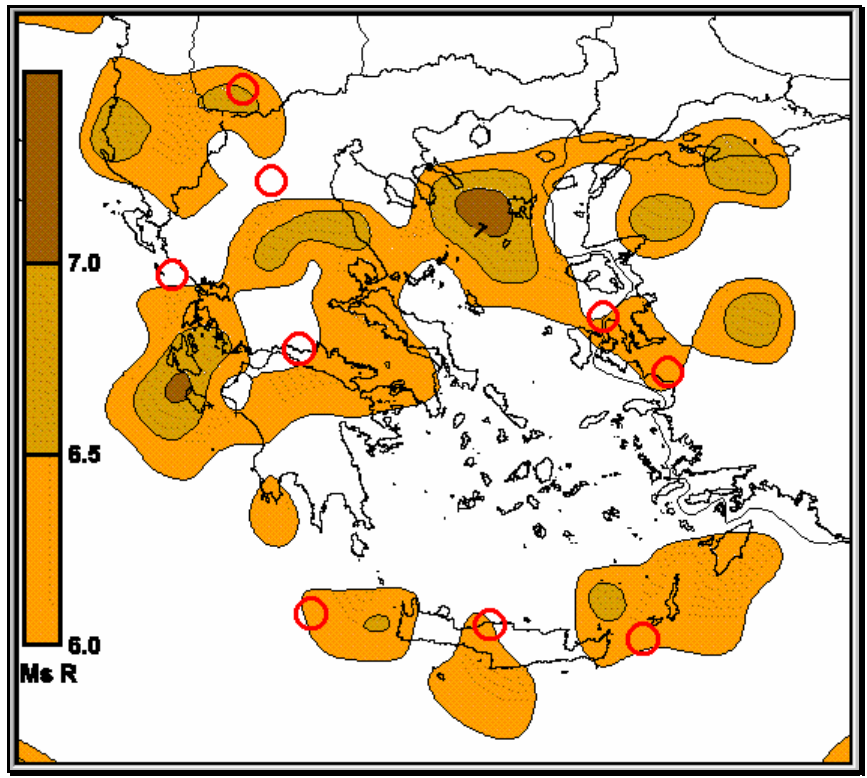

Fig. 22. Strong seismic events (red circles, Ms>=6R) during the period 1990 – 1995.
**EQtot = 9, EQin = 7, P = EQin / EQtot = 77.7%**

**Period: 1995 - 2000**

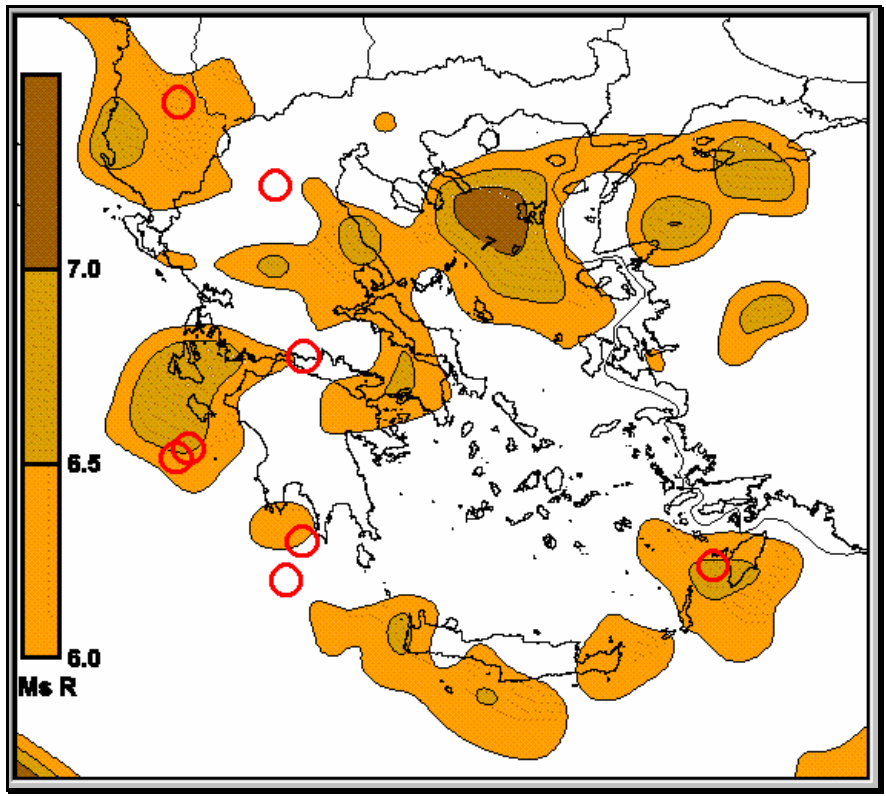

Fig. 23. Strong seismic events (red circles, Ms>=6R) during the period 1995 – 2000.
**EQtot = 8, EQin = 6, P = EQin / EQtot = 75%**

**Period: 2000 - 2003**

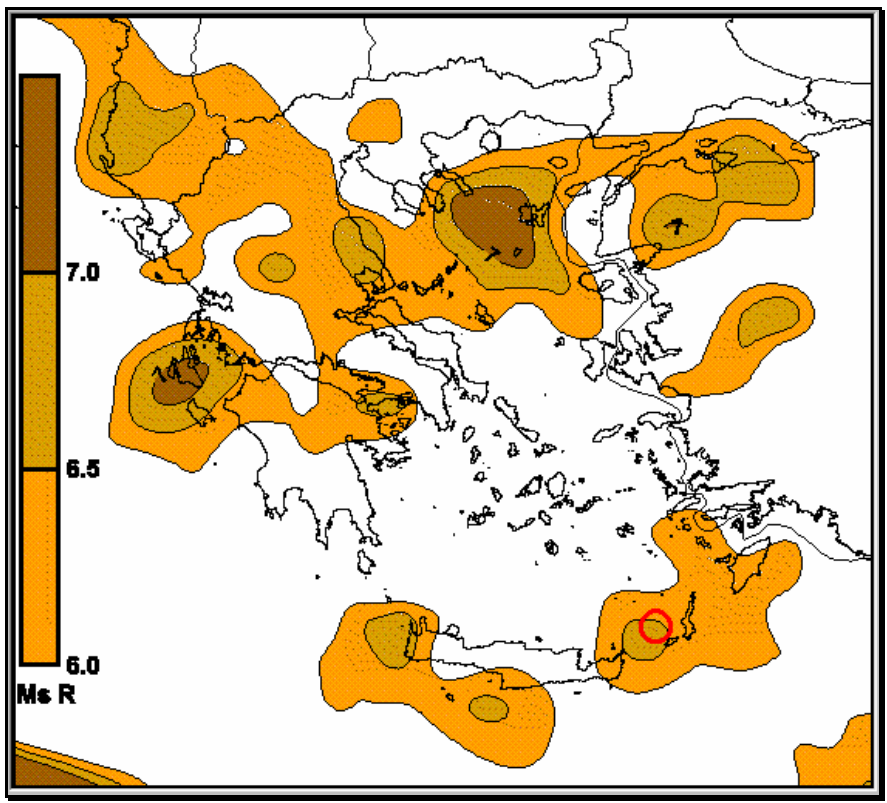

Fig. 24. Strong seismic events (red circles, Ms>=6R) during the period 2000 – 2003.

For the last period (2000 – 2003), the **(P)** value had not been calculated, since the seismic activity was still in progress (Thanassoulas et al., 2003) and two more years had to elapse. This map was compiled on 2003, when this work was presented for the first time. A refreshed map (compiled for the year 2005) will be presented and discussed in the subsequent text.

## 4. Discussion – Conclusions.

The Greek territory, considered as a unit area, appears to be continuously charged with seismic energy, since 1950. This is the result of the application of the energy flow model that indicates a continuous decrease, in total, of the seismic energy release in the Greek area. The later, is demonstrated, in figure **(2)**, where the energy flow rate **(EFL)** values decrease, along time, since 1950 and consequently, excess seismic energy is stored in the lithosphere. Therefore, it is expected that, some time in the future, the stored, seismic energy will be released through the occurrence of some strong seismic events (Thanassoulas et al., 2003).

The detailed study of the prepared maps **(fig. 4 – 17)**, for consecutive periods of 5 years (1970 – 2000), reveals the dynamic character of the seismic potential / maximum expected magnitude. A significant increase of seismic, potential charge is observed even within a five years period.

The observed, increased values of the seismic potential, in all the prepared maps, are distributed along the axis of the North Anatolian Fault Zone and the Southern Aegean Seismic Arc. The later, complies with the seismological observations which concern the spatial distribution of the strong seismic events in the Greek territory.

The periods of time, when the seismic potential has reached large levels, is indicated by the presence of expected earthquake magnitudes levels larger than seven **(7R)**. A very good example of the later is demonstrated in figure **(20)** for the period 1980 – 1985. The compiled map for the year 1980 indicates the presence of three centers of large seismic potential accumulation (Ms >7R, Kefalonia island, Korinthos – Alkyonides – Thiva area, Limnos island). In the next five years (1980 – 1985) that followed, 17 strong (M>6R) seismic events occurred, releasing a significant amount of seismic energy. The later is made clear by the next compiled map **(fig. 21, 1985)** and the small number of strong seismic events that occurred (only 3) in the following five years period (1985 – 1990).

What is more important, as a result of the comparison of the maps to each other, is the fact that a period of large seismic activity discharges the Greek territory for only .5 – 1.0 R. Therefore, distinct areas can be considered as being at a state of continuous, large seismic charge (expected EQ levels up to 6R) and slight, occasional increases of stress, trigger strong seismic events in the same areas.

In all these "prognostic" maps, which were compiled for the years 1970, 1975, 1980, 1985, 1990, 1995, 2000, were superimposed the corresponding large (Ms>6R) EQs for the next five years period of time. The percentage of success ranges from 62.5%, for the period 1975 – 1980, to a value of 88.2%, which was achieved for the period 1980 – 1985, in a total of 17 strong EQS, which occurred during this period. Obtained values of 100% for the percentage of success are based in a small number of strong seismic events and could be of no statistical value at all. Nevertheless, these seismic events still verify the validity of the method, used for the calculation of these seismic, potential maps.

As far as it concerns the small number of the seismic, strong events that didn't occurred in the predefined areas, it could be attributed to errors in compilation of these maps. A major error can be introduced by inadequate, available seismic data for the calculation of the cumulative energy graph **(Ec)** vs. time, for each used frame, during the application of the lithospheric energy flow model.

The compiled seismic potential maps, for different periods of time, were compared with the Athens EQ (1999, Ms = 5.9R), Karpathos (2001, Ms = 6.6R) and Zakynthos-Kefalonia EQ (2003, Ms = 5.8R) strong seismic events in Greece.

In the following figure **(25),** the seismic potential map of Greece, for the year 1995, is shown, along with the strong (Ms >6R) seismic events that occurred in the period 1995 - 2000. The red arrow indicates the place where the Athens EQ occurred (Ms = 5.9R). It coincides with the narrow area, of expected events of Ms >6.5R, observed, in the same place. A comparison of this map with the one in figure **(22),** compiled for the year 1990, indicates that seismic potential at the regional area of Athens was built-up during the time span of 1990 – 1995.

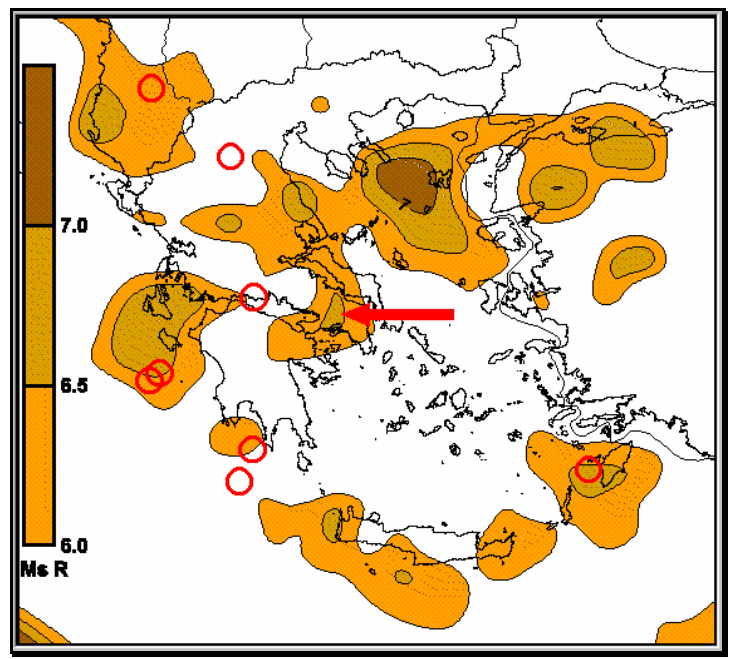

Fig. 25. Seismic potential spatial distribution in Greece is presented for the year 1995. The red arrow indicates the location of Athens (1999) EQ.

A comparison of figure **(25)** with figure **(24)**, of the seismic potential for the year 2000, reveals that there is still accumulated, large seismic potential, at a rather short distance, west of Athens.

The next example concerns Karpathos strong EQ (2001, Ms = 6.6R). The corresponding map of the seismic potential was compiled for the year 2000. The red arrow indicates the location of the earthquake. The coincidence of its epicenter with the area of increased, seismic potential (expected EQ with Ms >6.5R) is more than evident.

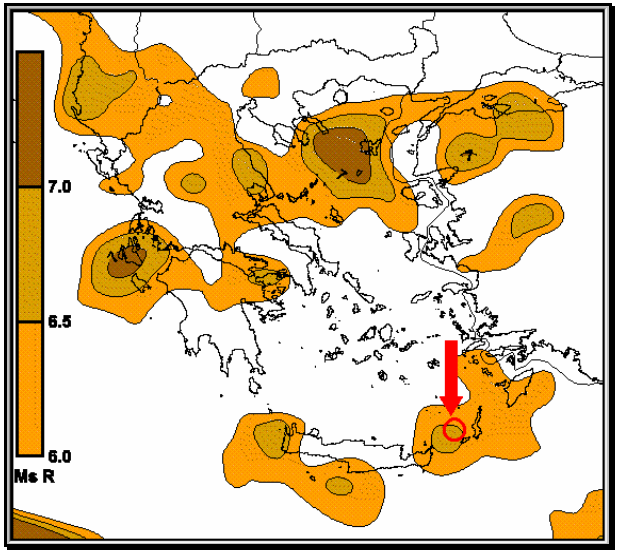

Fig. 26. Seismic, potential spatial distribution, in Greece is presented, for the year 2000. The red arrow indicates the location of Karpathos (2001) EQ.

The last example refers to Zakynthos - Kefalonia (2003, Ms = 5.8R). The red arrow indicates the location of this EQ. Even if this EQ is not of a magnitude larger than 6R, it is still considered as a rather strong one and of significant magnitude since it occurred near inhabited areas.

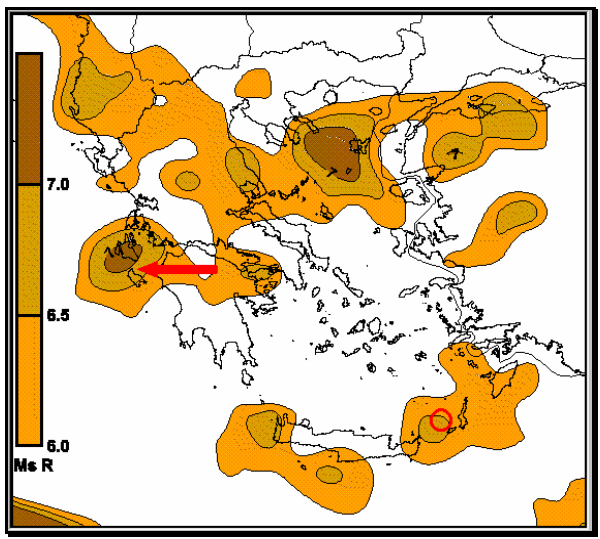

Fig. 27. Seismic, potential spatial distribution in Greece is shown for the year 2000. The red arrow indicates the location of Zakynthos - Kefalonia (2003) EQ.

Apart from the previous, presented examples, a statistical study was applied on the obtained results, as follows:

a.  the total number of strong seismic events, during the period 1970 – 2000, is equal to 50.

b.  the total number of the seismic events that occurred, in the predefined as seismically active areas, is 41.

c.  for the total period between 1970 – 2000 the **P** value is calculated as:

**P = 41/50 = 82%**

For the last compiled map **(fig. 24)**, for the year 2000, it was not possible to calculate (on 2003) a **P** value, since two more years were left for completing the five years time interval, needed. Therefore, only a statistical extrapolation, as follows, was made, based, on the previous results (Thanassoulas et al. 2003).

Since only one strong earthquake occurred in the period 2000 – 2003, and assuming the value of 82% is the correct one, **then 5 more strong EQs are expected,** within the next two years, four of them must occur in the predefined areas, so that the P value is: **P = 5/6 = 83%** that is very close to the calculated 82% (average value, calculated, for the entire time period of the study), and it is the result of the <u>smallest pair of integer numbers found</u>.

If a different scenario is followed, for a minimum **P** value found of 62.5% (period 1975 – 1980), then, by following the above procedure, it is found that **four more strong EQs must occur**, two of them in the predefined area, so that the **P** value is: **P = 3/5 = 60%** being very close to the assumed value of 62.5%. Regardless the used statistical hypothesis, 4 – 5 strong (Ms > 6R) seismic events were, theoretically, expected, to occur within the next period of time up to 2005, included (Thanassoulas et al. 2003).

What was presented so far was the analysis of the data available up to 2003. Since five years have elapsed, from the time (2003) when the later scenarios had been presented, it is now possible to present the entire 2000 – 2005 period seismic potential map and to compare it with the "guesses", made on 2003. This is shown in the following figure **(28)**.

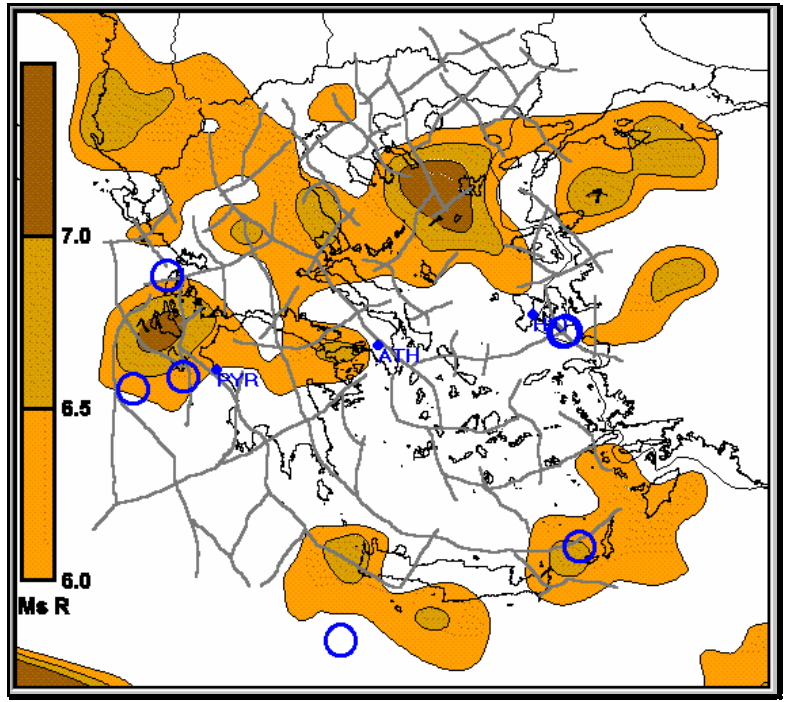

Fig. 28. Strong seismic events (blue circles, Ms>=6R) during the period of time 2000 – 2005.

The number of strong (M > 6R) seismic events that took place is seven (7) and four (4) of them are located in seismically charged areas. The calculated **P** value is:

**P = 4/7**     or     **57%**

If the double, seismic event of Hios Island (eastern Greece), is considered <u>as of marginal success</u>, then the P value becomes:

**P = 6/7**     or     **86%**

Actually, both hypothetical scenarios were superseded by a larger number of earthquakes **(6)** verifying thus the number of expected strong EQs in the period 2000 – 2005.

It is worth to compare the former static, seismic hazard map **(fig. 1)**, compiled by the seismological data only, with the seismic, potential maps, compiled, by the application of the energy flow model of the lithosphere. The later indicates, in medium term periods of time, the dynamic change of the seismic potential charge of the lithosphere and therefore, it can be used so that medium term measures against any seismic event, can be taken, by the State Authorities, in areas prone to intense, seismic activity. Consequently, the seismic hazard map of Greece (**fig. 1**) should be used along with the compiled seismic potential maps. The later should be recompiled frequently, since the current seismic potential changes dynamically, in medium time intervals, along the time span.

Finally, using the earthquake data available up to 2008, an update is made of the figure (**2**) and is presented in the subsequent figure (**29**) aside of figure (**2**).

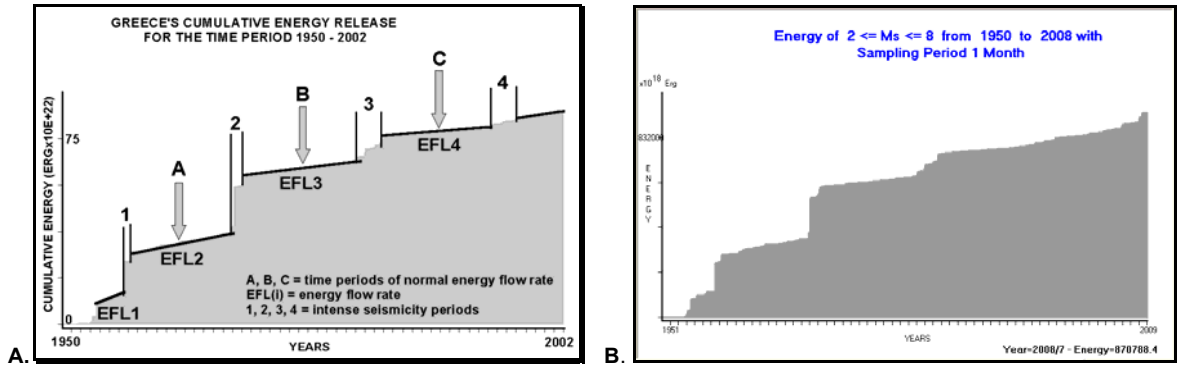

Fig. 29. Greece's cumulative energy release is presented for the time period 1950 – 2003 (**A**) and 1950 – 2008 (**B**).

The comparison of figure (**29, B**) to the one of figure (**29, A**) suggests that the entire Greek territory is being subjected to accelerating deformation for the last 5 - 6 years. This is more evident after enlarging the last 8 years (1999 – 2008) of the graph (**29, B**). The later is presented in the subsequent graph of figure (**30**).

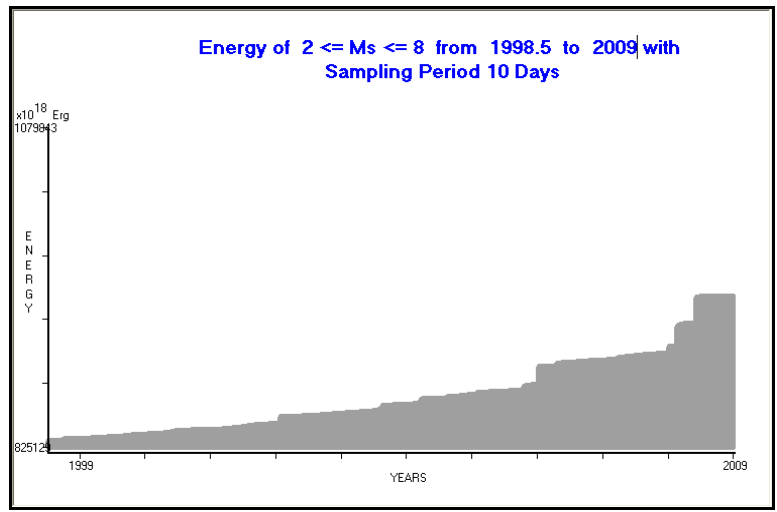

Fig. 30. Enlargement of right end part figure (**29, B**)

The accelerating deformation mode of figure (**30**) is enhanced by fitting a 6$^{th}$ degree time function presented in the following figure (**31**).

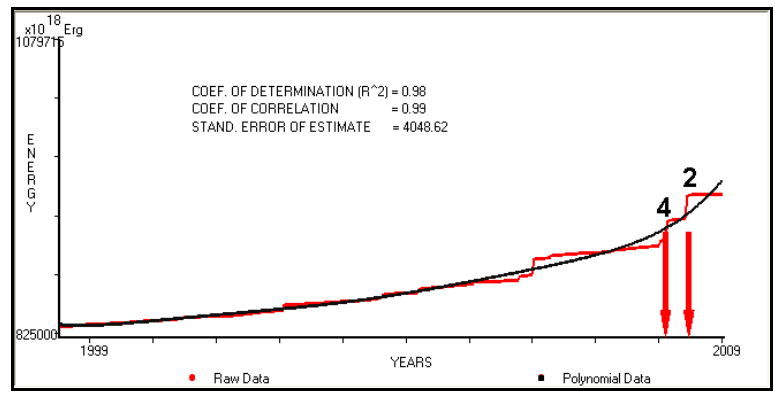

Fig. 31. Time function of 6$^{th}$ order polynomial (black line) calculated from the raw cumulative seismic energy release data (red line) of figure (**30**). The two red arrows indicate multiple (4 and 2, M>6R) large seismic events in 2008.

The two sudden steps in 2008, observed in the cumulative seismic energy release graph, correspond to the intense seismic activity which took place just recently. The first red arrow corresponds to a group of 4 earthquakes (M>6R) which took place in January and February of 2008, while the second red arrow corresponds to a group of 2 earthquakes (M>6R) which took place on June 2008.

A more indicative figure, referring to the accelerating deformation of the same time period, is presented by calculating the time - grad of the previously calculated 6$^{th}$ order time polynomial function. This operation is presented in the subsequent figure (**32**).

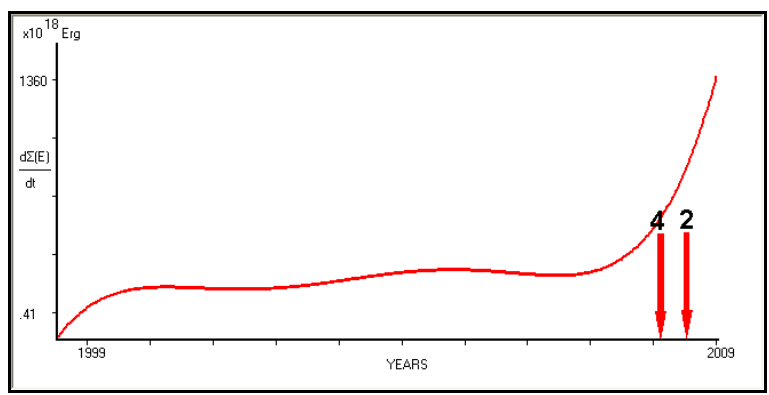

Fig. 32. Time - grad calculated on the analytical expression of the 6$^{th}$ order time polynomial function. The two red arrows indicate multiple (4 and 2, M>6R) large seismic events in 2008.

Hence, the statement made (on 2003 and pointed in this work) in section (**2**) as: "for the last 52 years, Greece, was charged continuously with seismic strain energy that will be released some time in the future, at some seismically prone areas" was validated quite faster than what it was expected to happen when it was postulated on 2003.

## 5. References.

Bowman, D.D., Ouillon, G., Samnis, C.G., Sornette, A., and Sornette, D., 1998. An observational test of the critical earthquake concept., Journal of Geophysical Research, Vol. 103, No. B10, pp 24,359-24,373, October 10.
Bufe, C.G., Varnes, D.J., 1993. Predicted modeling of the seismic cycle of the greater San Francisco Bay region, J. Geophys. Res., 98, 9871- 9883.
Di Giovambattista, R., Tyupkin, Y., 2001. An analysis of the process of acceleration of seismic energy emission in laboratory experiments on destruction of rocks and before strong earthquakes on Kamchatka and in Italy., Tectonophysics, 338, pp. 339-351.
Main, I., 1995. Earthquakes as critical phenomena: implications for the probabilistic seismic hazard analysis. Bull. Seism. Soc. Am. 85, 1299 -1308.
Main, I., 1996. Statistical physics, seismogenesis and seismic hazard, Rev. Geophys., 34(4), 433-462.
Makropoulos, K.C., and Burton, P.W., 1985. Seismic hazard in Greece, II, Ground acceleration, Tectonophysics, 117, 259-294.
Papazachos, B.C, Kiratzi, A.A., Hatzidimitriou, P.M., Papaioannou, C.A., and Theodoulidis, N.P., 1985. Regionalization of seismic hazard in Greece. Proc. 12$^{th}$ Regional Seminar on Earthquake Engineering, EAEE – EPPO, Halkidiki, Greece, September 1985.
Papazachos, B.C., 1989. Seismicity of the Aegean and surrounding area. "Publ. Geophys. Lab. Un. of Thessaloniki", 4, pp. 1-27.
Papazachos, B,K., Makropoulos, K., Latousakis, I., Theodoulidis, N., 1989. Compilation of seismic hazard map of Greece., 2nd report for Organization for Seismic Plannining of Greece (OASP).
Papazachos, B., and Papazachos, C., 2000. Accelerated Preschock Deformation of Broad Regions in the Aegean Area, Pure Appl. Geophys, 157, 1663, 1681.
Papazachos, B., Karakaisis, G.F., Papazachos, C., Scordilis, E., Savaidis, A., 2001. A method for estimating the origin time of an ensuing mainshock by observations of preshock crustal seismic deformation, Bul. Geol. Soc. Greece, XXXIV/4, 1573-1579.
Papazachos, C.B., and Papazachos, B.C., 2001a. Precursory accelerated Benioff strain in the Aegean area., Ann. Geofisica, 144, 461-474.
Papazachos, C.B., Karakaisis, G.F., Savvaidis, A.S., and Papazachos, B.C., 2002. Accelarating seismic crustal deformation in the Southern Aegean area., Bull. Seism. Soc. Am., 92, 570-580.
Thanassoulas, C., 2007. Short-term Earthquake Prediction, H. Dounias & Co, Athens, Greece. ISBN No: 978-960-930268-5
Thanassoulas, C. 2008a. The seismogenic area in the lithosphere considered as an "Open Physical System". Its implications on some seismological aspects. Part – I. Accelerated deformation. arXiv.org:0806.4772v1 [physics.geo-ph]
Thanassoulas, C. 2008b. The seismogenic area in the lithosphere considered as an "Open Physical System". Its implications on some seismological aspects. Part – II. Maximum expected magnitude. arxiv.org:0807.0897v1 [physics.geo-ph]
Thanassoulas, C., Klentos, V., 2001. The "energy-flow model" of the earth's lithosphere. Its application on the prediction of the "magnitude" of an imminent large earthquake. The "third paper". IGME, Open file report: A.4384, Greece.
Thanassoulas, C., Klentos, V. 2003. Seismic potential map of Greece, calculated by the application of the "Lithospheric energy flow model"., IGME, Open File Report A. 4402, Athens, Greece, pp. 1-25.
Tselentis, G., 1997. Contemporary Seismology, Seismic Energy, Vol. 2, pp. 511-514.
Tzanis, A., and Vallianatos, F., 2003. Distributed power-low seismicity changes and crustal deformation in the SW Hellenic ARC., Natural Hazards and Earth System Sciences, 3, 1-17.
Varnes, D.J., 1987a. Foreshock seismic energy release functions – tools for estimating time of main shocks (abstract). Seismol. Res. Lett. 58, 21.
Varnes, D.J., 1987b. Foreshock seismic energy release functions – tools for estimating time of main shocks. USGS. Open File Report 87- 429, 1- 44.
Varnes, D., 1989. Predicting earthquakes by analyzing accelerating precursory seismic activity., Pageoph., 130, 4, pp. 661-686.

14